\documentclass[graybox]{svmult}

\usepackage{type1cm}        % activate if the above 3 fonts are
\usepackage{makeidx}         % allows index generation
\usepackage{graphicx}        % standard LaTeX graphics tool
\usepackage{multicol}        % used for the two-column index
\usepackage[bottom]{footmisc}% places footnotes at page bottom

\usepackage{newtxtext}       % 
\usepackage{newtxmath}       % selects Times Roman as basic font
\usepackage{subfigure}
\makeindex             % used for the subject index
\begin{document}

\title*{Asteroseismic stellar modelling: systematics from the treatment of the initial helium abundance}
\titlerunning{Systematics from the treatment of the initial helium abundance} %for an abbreviated version of
\author{
Nuno Moedas, Benard Nsamba, and Miguel T. Clara}
\authorrunning{N. Moedas et al.} %for an abbreviated version of
\institute{
Nuno Moedas and Miguel T. Clara
\at Instituto de Astrof\'{\i}sica e Ci\^{e}ncias do Espa\c{c}o, Universidade do Porto,  Rua das Estrelas, PT4150-762 Porto, Portugal
\at Departamento de F\'{\i}sica e Astronomia, Faculdade de Ci\^{e}ncias da Universidade do Porto, PT4169-007 Porto, Portugal\\
\email{nunomoedas17@gmail.com}
%%%%%%%%
\and Benard Nsamba 
\at Max-Planck-Institut f\"{u}r Astrophysik, Karl-Schwarzschild-Str. 1, D-85748 Garching, Germany
\at Instituto de Astrof\'{\i}sica e Ci\^{e}ncias do Espa\c{c}o, Universidade do Porto,  Rua das Estrelas, PT4150-762 Porto, Portugal \\
\email{nsamba@mpa-garching.mpg.de} 
%%%%%%%%
%\and Tiago L. Campante
%\at Instituto de Astrof\'{\i}sica e Ci\^{e}ncias do Espa\c{c}o, Universidade do Porto,  Rua das Estrelas, PT4150-762 Porto, Portugal
%\at Departamento de F\'{\i}sica e Astronomia, Faculdade de Ci\^{e}ncias da Universidade do Porto, PT4169-007 Porto, Portugal
%%%%%%%%
%\and Margarida S. Cunha
%\at Instituto de Astrof\'{\i}sica e Ci\^{e}ncias do Espa\c{c}o, Universidade do Porto,  Rua das Estrelas, PT4150-762 Porto, Portugal
%%%%%%%%
%\and M\'{a}rio J. P. F. G. Monteiro
%\at Instituto de Astrof\'{\i}sica e Ci\^{e}ncias do Espa\c{c}o, Universidade do Porto,  Rua das Estrelas, PT4150-762 Porto, Portugal
%\at Departamento de F\'{\i}sica e Astronomia, Faculdade de Ci\^{e}ncias da Universidade do Porto, PT4169-007 Porto, Portugal
%%%%%%%
%%%%%%%%
%\and S\'{e}rgio G. Sousa
%\at Instituto de Astrof\'{\i}sica e Ci\^{e}ncias do Espa\c{c}o, Universidade do Porto,  Rua das Estrelas, PT4150-762 Porto, Portugal\\
\\
%\vspace{70pt}
}
\maketitle
%%%%%%%%%%%%%%%%%%%%%%%%%%%%%%%%%%%%%%%%%%%%%%%
\abstract
{Despite the fact that the initial helium abundance is an essential ingredient in modelling solar-type stars, its abundance in these stars remains a poorly constrained observational property. This is because the effective temperature in these stars is not high enough to allow helium ionization, not allowing any conclusions on its abundance when spectroscopic techniques are employed.
To this end, stellar modellers resort to estimating the initial helium abundance via a semi-empirical helium-to-heavy element ratio, anchored to the the standard Big Bang nucleosynthesis value.
Depending on the choice of solar composition used in stellar model computations, the helium-to-heavy element ratio, ($\Delta Y/\Delta Z$) is found to vary between 1 and 3.
In this study, we use the {\it Kepler} LEGACY stellar sample, for which precise seismic data is available, and explore the systematic uncertainties on the inferred stellar parameters (radius, mass, and age) arising from adopting different values of
$\Delta Y/\Delta Z$, specifically, 1.4 and 2.0.
The stellar grid constructed with a higher $\Delta Y / \Delta Z$ value yields lower  radius and mass estimates. 
We found systematic uncertainties of 1.1\%,  2.6\%, and 13.1\% on radius, mass, and ages, respectively. 
%We also report a good agreement in terms of the derived mean densities, with statistical uncertainties comparable to the systematic uncertainties.
%By comparing our results with the ESA's PLATO scientific requirements, we show that differences in the treatment of $\Delta Y/\Delta Z$ lead to results in the limits imposed (being these 2-4\% for radius, 10-15\% for mass, and ~10\% for age).
}

%\twocolumn
%%%%%%%%%%%%%%%%%%%%%%%%%%%%%%%%%%%%%%%%%%%%%%
\section{Introduction}
Chemical abundances are some of the most essential ingredients in stellar modelling, complementing our understanding of the formation, structure, and evolution of stars. Solar abundances are commonly adopted in stellar evolution codes, e.g., Modules for Experiments in Stellar Astrophysics (MESA; \cite{Pax}), 
among others, however, some significant discrepancies exist. 
Among the different element abundances, helium abundance measurements in solar-type stars are one of the most poorly constrained ingredients in stellar modelling. This is because the temperature required to excite an atomic transition of the helium exceeds 20,000K, a value higher than the characteristic effective temperature of the solar-type stars (e.g., \cite{Gennaro}).

To overcome the helium abundance problem when constructing stellar models, 
a common solution is use the ``Galactic chemical evolution law'' in that the iron content ([Fe/H]) is transformed into fractional abundances (i.e., hydrogen mass fraction, $X$, helium mass fraction, $Y$, and heavy elements mass fraction, $Z$) via the helium enrichment ratio ($\Delta Y / \Delta Z$) using the expression
\begin{equation}
\frac{\Delta Y}{\Delta Z} = \frac{Y - Y_0}{Z - Z_0}
 \label{helium}
\end{equation}
set to the BNN values of $Z_{0}$= 0.0 and $Y_{0}$= 0.2484 \cite{Cyburt}.
\cite{Jimenez} reported $\Delta Y / \Delta Z $ = 2.1 $\pm$ 0.4 using observations of nearby K dwarfs and a set of isochrones. Similar results were found by \cite{Casagrande} using a set of Padova isochrones and observations of nearby K dwarfs (i.e., $\Delta Y / \Delta Z $ = 2.1 $\pm$ 0.9). \cite{Balser} published  $\Delta Y / \Delta Z $ = 1.6 obtained using metal-poor H II regions, Magellanic cloud H II regions and M17 abundances, while taking into account the effects of temperature fluctuations. Interestingly, when using only galaxy H II region S206 and M17, \cite{Balser} determines $\Delta Y / \Delta Z $ = 1.41 $\pm$ 0.62, a value reported to be consistent with that from standard chemical evolution models. Depending on the choice of solar composition, \cite{Sereneli} reported the initial helium abundance of the Sun to be in agreement with a slope of 1.7 $\leq \Delta Y / \Delta Z \leq$ 2.2. In general, acceptable values for the helium enrichment ratio tend to vary between 1 and 3.

\cite{Lebreton} reported a scatter of $\sim$ 5\% in mass arising from the treatment of  initial helium mass fraction. Using synthetic data of about 10,000 artificial stars, \cite{Valle} found a systematic bias of 2.3\% and 1.1\% in mass and radius, respectively, arising from a variation of $\pm 1$ in $\Delta Y / \Delta Z$. Further, \cite{ValleB} reported a systematic bias in age to be about one-fourth of the statistical error in the first 30\% of the evolution, while its negligible for more evolved stars. 
The treatment of the initial helium mass fraction in stellar models is therefore a substantial source of systematic uncertainties on stellar properties (such as mass, radius, and age) derived through forward modelling techniques.

In this article, we take advantage of the available stars with multi-year {\it {Kepler}} photometry \cite{Lund} and with parallax measurements from the Gaia mission. 
We complement all these constraints with spectroscopic constraints (i.e., effective temperature and metallicity) and quantify systematic uncertainties  on stellar parameters (mainly radius, mass, and age) arising from the variation in $\Delta Y / \Delta Z$ values used Equation~\ref{helium}.
%In addiction, we compare our results with the ESA's PLATO\footnote{European Space Agency and PLAnetary Transits and Oscillations of stars.} scientific requirements, that present accuracy limits of 10-15\% for the mass, 2-4\% for the radius, and 10\% for the stellar age \citep{PLATO}.

%%%%%%%%%%%%%%%%%%%%%%%%%%%%%%%%%%%%%%%%%%%%%%%%%%%
\section{Target Sample}
\begin{figure}[b]
    \begin{minipage}{0.35\columnwidth}
    \caption{$\Delta\nu$ -- T$_{\rm eff}$ diagram. Each circle represents a target star colour coded according to its metallicity. The black lines represent stellar evolutionary tracks, ranging in mass from 0.8M$_\odot$ to 1.5M$_\odot$, with $Z=0.02$ and a mixing length parameter ($\alpha_{\rm MLT}$) of 1.8 constructed using MESA.}
    \end{minipage}
    \hspace{.07\columnwidth}
    \begin{minipage}{0.57\columnwidth}
    \begin{center}
        \includegraphics[width=1\columnwidth]{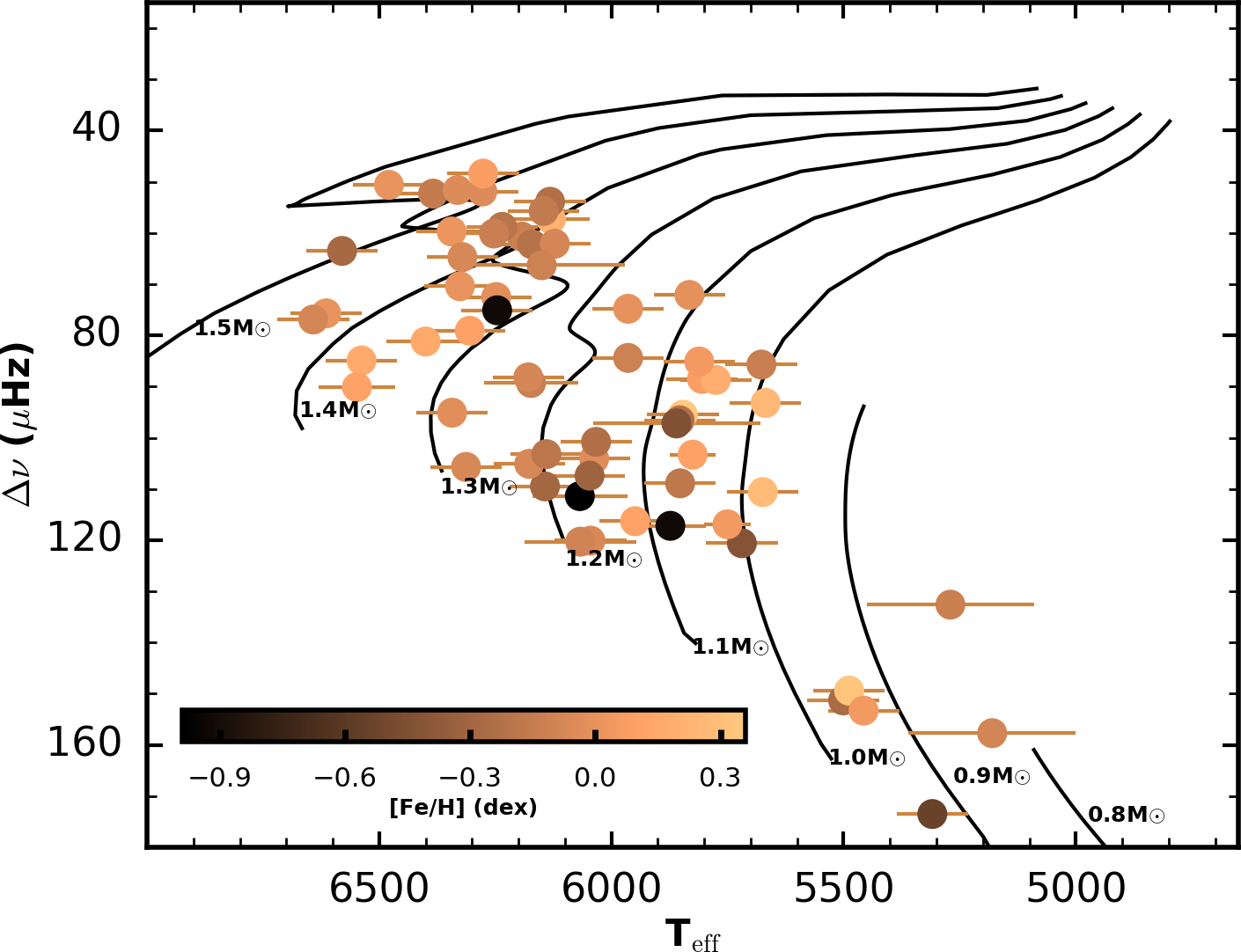}
    \par\end{center}
    \end{minipage}
    \label{tracks}
\end{figure}

%\label{sample}
%We describe the {\it Kepler} data used as our target sample in Sect.~\ref{target}, while the characteristics of our stellar grids and the optimisation tool used in the forward modelling routines are discussed in Sect.~\ref{models}.
%
%%%%%%%%%%%%%%%%%%%%%%%%%%%%%%%%%%%%%%%%%%%%%%%%%%%%%%%%%
%\subsection{ \textit{Kepler} data}
\label{target}
%%%%%%%%%%%%%%%
Our sample consists of  66 {\it Kepler} LEGACY stars \cite{Silva,Lund} with at least 12 months of short cadence data ($\Delta t$ = 58.89 s). Fig.~\ref{tracks} shows the location of our sample on a  $\Delta \nu$ (large separation) -- T$_{\rm eff}$ diagram. The spectroscopic parameters (metallicity, [Fe/H], and effective temperature, $T_{\rm eff}$) for each star were obtained from  \cite{Silva} and references therein. 
%%%%%%%%%%%%%%%%%%%%%%%%%%%%%%%%%%%%%%%%%%%
%%%%%%%%%%%%%%%%%%%%%%%%%%%%%%%%%%%%%%%%%%%%%%%%%%%
\cite{Travis} used {\it Gaia} Data Release 2 ({\it Gaia} DR2) parallaxes as inputs in the stellar classification code ``isoclassify'' \cite{Huber} to derive stellar radii for majority of the {\it Kepler} stars. Adopting the stellar radii and $T_{\rm eff}$ measurements in the Stefan-Boltzmann relation,
we derived the stellar luminosities for the majority of our sample.
For the stars in our sample that were not analysed by \cite{Travis}, we obtained their luminosities using the expression \cite{Pijpers}
\begin{equation}
\log(L/{\rm L}_\odot)=4.0+0.4M_{\rm bol,\odot}-2.0\log(\pi[{\rm mas}])-0.4(V-A_V+BC_V),
\label{LvsExt}
\end{equation}
\vspace{-1pt}
where  $M_{\rm bol,\odot}$ is the bolometric magnitude of the Sun with a value of $4.73$ mag, $\pi[\rm mas]$ is the parallax, $V$ is the magnitude in the $V$ band obtained from \cite{Hog},  $A_V$ is the extinction in the $V$ band taken from \cite{Mathur}, and $BC_V$ is the bolometric correction calculated using the polynomial expression from \cite{Torres}. We note that for the binary system 16 Cyg, we used the luminosities presented by  \cite{Metcalfe}.

%%%%%%%%%%%%%%%
\vspace{-13pt}
\subsection{Stellar models}
\label{models}
%%%%%%%%%%%%%%
We built two stellar grids (namely A and B) varying mainly in the treatment of initial helium mass fraction ($Y$) using MESA version 9793.  The evolution tracks were only terminated when models reach:
(i) a stellar age of $\sim$ 16 Gyr and (ii) a point along the evolutionary tracks where the log$(\rho_c)$ = 4.5 ($\rho_c$ is the central density). We note that only models from the Zero Age Main-Sequence (ZAMS) to the termination point  were stored.

\begin{table}
\centering 
\caption{Stellar grid constituents.} 
\begin{tabular}{p{1cm}p{1.5cm}p{1.5cm}p{1.5cm}p{1.5cm}p{0cm}}       
\hline\noalign{\smallskip} % inserts single horizontal line
Grid &	Mass (M$_\odot$)	&  Diffusion & Overshoot & $\Delta Y$ $/$ $\Delta Z$\\
\noalign{\smallskip}\svhline\noalign{\smallskip}
 A 	& 0.7 -- 1.1				&	Yes			&	No	&  1.4\\
	& 1.2 -- 1.6	    		&	No			&	Yes &  1.4 \\
\hline\noalign{\smallskip}
 B 	& 0.7 -- 1.1		    &	Yes			&	No	&  2.0\\
    & 1.2 -- 1.6			&	No			&	Yes	&   2.0\\
\hline 
\end{tabular}
\label{tab_grid}
% title of Table
\end{table}

Table~\ref{tab_grid} summarizes the different grid constituents. The evolution tracks vary in mass M $\in$ [0.7 -- 1.6] M$_\odot$ in steps of 0.05 M$_\odot$, Z $\in$ [0.004 -- 0.04] in steps of 0.002, and $\alpha_{\rm mlt}$ $\in$ [1.0 -- 3.0] in steps of 0.4.
Atomic diffusion is known to be an efficient transport process in low mass stars and was included in our low mass models (see Table~\ref{tab_grid}) based the description of \cite{Thoul}. 
For models with M $\in$ [1.2 -- 1.6] M$_\odot$, we include convective core overshoot by  adopting the exponential diffusive overshoot procedure as implemented in MESA based on \cite{Herwig}.
%\vspace{-5pt}
%\begin{equation}
%\rm Q_{\rm c} = \rm Q_{\rm 0} ~ exp\left(   \frac{-2d}{\rm f \cdot H_{\rm p}}    \right) %~~,
%\label{rad}
%\vspace{-5pt}
%\end{equation}
%where $\rm Q_{\rm c}$ is the diffusion coefficient in the overshoot region, $\rm Q_{\rm 0}$ is the diffusion coefficient in the convectively unstable region near the convective boundary determined using the mixing length theory (MLT), $\rm H_{\rm p}$ is the pressure scale height, and $\rm d$ is the distance from the edge of the convective zone.
The overshoot parameter was set to vary in the range [0.0 -- 0.03] in steps of 0.005. We note models in the mass range [1.1 -- 1.15] M$_\odot$ lie in the transition region where models may develop convective cores. For models within this mass range with a convective core, both diffusion and  core overshoot were included.

%\citet{Nsambaa} has demonstrated that some models with masses between 1.1 and 1.15 M$_\odot$ develop a convective core. Therefore, we included both diffusion and convective core overshoot in such models, while only diffusion was included for those with radiative cores.
The general input physics used for the grids include nuclear reaction rates obtained from Joint Institute for Nuclear Astrophysics Reaction Library (JINA REACLIB; \cite{Cyburt}) version 2.2 with specific rates for $^{12}{\rm C}(\alpha,\gamma)^{16}{\rm O}$ and $^{14}{\rm N}(p,\gamma)^{15}{\rm O}$ described by \cite{Kunz} and \cite{Imbriani}, respectively. At high temperatures, OPAL tables  \cite{Iglesias} were used to cater for opacities while tables from \cite{Ferguson} were used at lower temperatures. Both grids A and B used the 2005 updated version of the OPAL equation of state \cite{Rogers}. The surface boundary of stellar models is described using the standard Grey-Eddington atmosphere. Both grids used metallicity mixtures from \cite{Grevesse}.We note that GYRE \cite{Townsend} was used to calculate model oscillation frequencies for spherical degrees $\ell=0,1,2,3$.

In order to generate a set of models that best-fit the seismic and classical constraints of our sample stars, we employ the Asteroseismic Inference on a Massive Scale (AIMS; \cite{Rendle}) yielding the mean and standard deviation of the posterior probability density functions of different stellar parameters.
%----------------------------------------------------------------------------------------------------
%%%%%%%%%%%%%%%%%%%%%%%%%%%%%%%%%%%%%%%%%%%

\begin{figure}[t]
    \includegraphics[width=0.49\columnwidth]{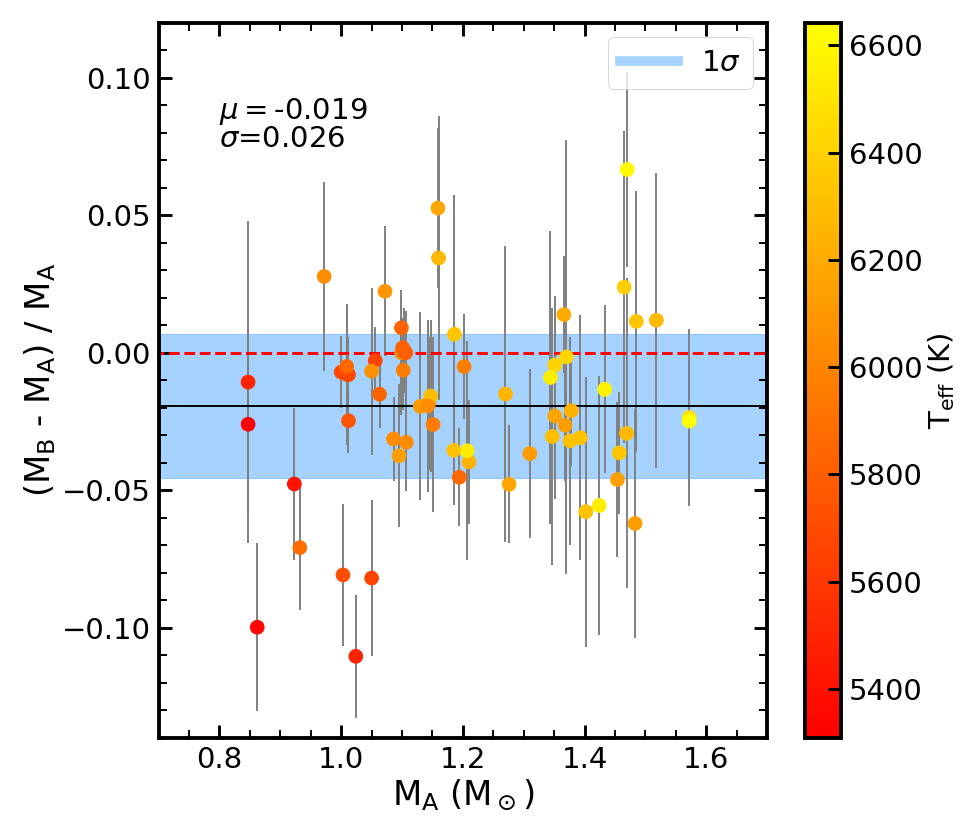}
    \includegraphics[width=0.49\columnwidth]{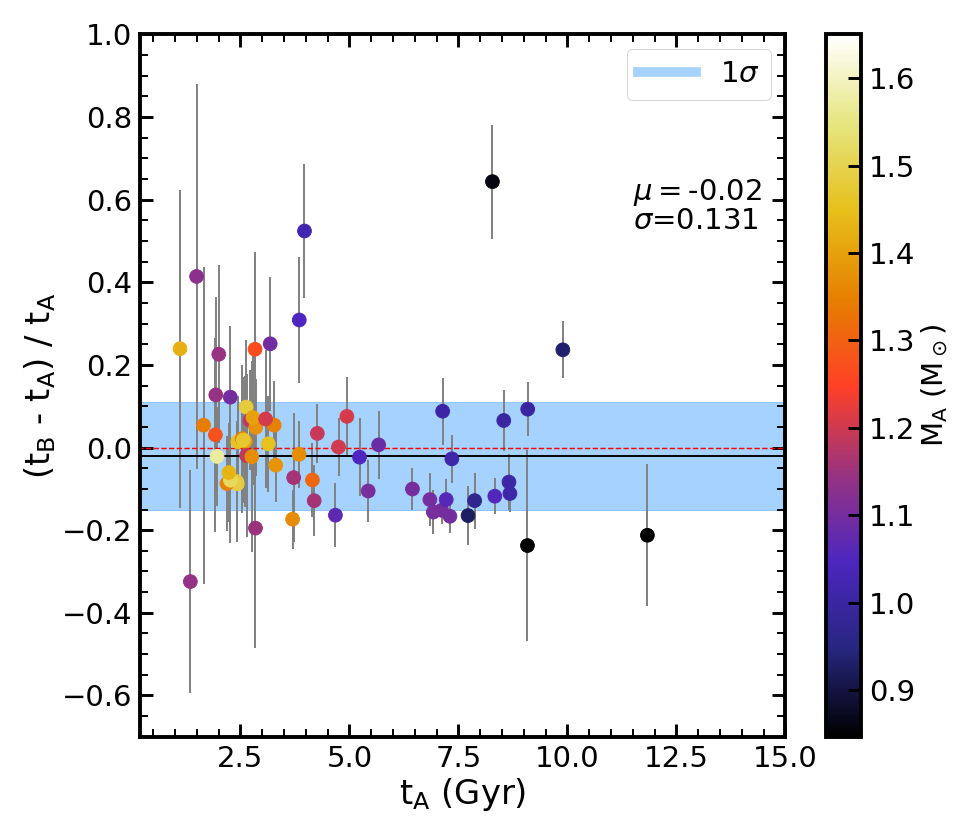}
    \begin{minipage}{0.49\columnwidth}
    \includegraphics[width=\columnwidth]{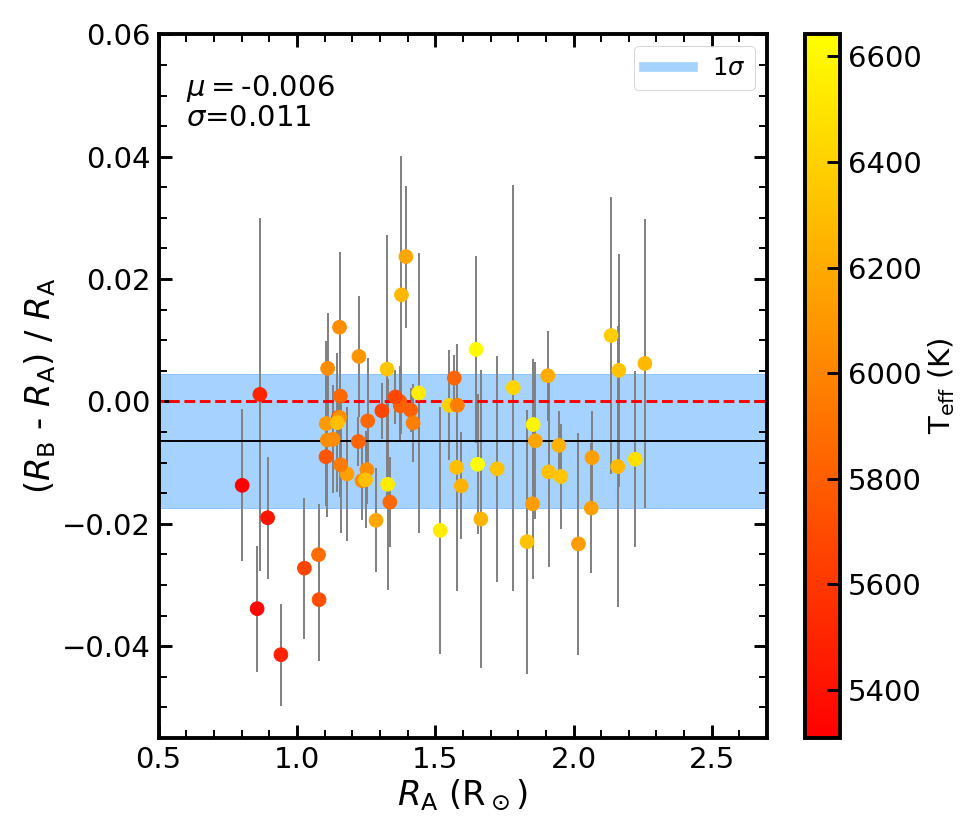}
    \end{minipage}
    \hspace{0.02\columnwidth}
    \begin{minipage}{0.45\columnwidth}
        \caption{Fractional differences in stellar mass, age, and radius resulting from the treatment of the initial helium mass fraction (abscissa values from grid A). The black line represents the bias ($\mu$), the scatter ($\sigma$) is represented by the blue  region, and the red line is the null offset.}
    \end{minipage}
    \hspace{0.02\columnwidth}
    \label{tracks11}
\end{figure}
%%%%%%%%%%%%%%%%%%%%%%%%%%%%%%%%%%%%%%%%%%%%%%%%%
\section{Discussion}
\label{results}
%%%%%%%%%%%%%%%%%%%%%%%%%%%%%%%%%%%%%%%%%%%%%%%%%%
%We will discuss in detail the results obtained from the comparison between the grids with different $\Delta Y/\Delta Z$ values.
The top left panel of Fig.~\ref{tracks11} shows that grid A yields optimal solutions with higher masses compared to grid B, with systematic uncertainties of $\sim$ 2.6\% and a bias of $\sim$ 2\%. A similar trend can be seen in the bottom panel of Fig.~\ref{tracks11}, with grid A yielding higher radii compared to grid B, with systematic uncertainties of $\sim$ 1.1\% and a bias of $\sim$ 0.6\%. 
Based on Equation~\ref{helium}, for a given value of $Z$, grid B (i.e., with $\Delta Y/\Delta Z$ = 2) has models with higher initial helium mass fraction ($Y$) compared to models in grid A (i.e., with $\Delta Y/\Delta Z$ = 1.4). 
This implies that models in grid B have a higher mean molecular weight which, in turn, increases the energy production rate resulting into an increase in the energy flux at the surface \textemdash~ increasing the model luminosity.

We stress that stellar luminosities of our target stars are included as part of the surface constraints in our optimisation process. That being said, in order to have best-fit models from grid A (i.e., with lower $\Delta Y/\Delta Z$ ratio) which satisfy the observed stellar luminosities, they should have higher masses and radius compared to those from grid B as shown in the top left and bottom panels of Fig.~\ref{tracks11}, respectively.
The top right panel of Fig.~\ref{tracks11} shows a relatively good agreement in the stellar ages from both grid A and B, with a bias ($\mu$) of 2\% and systematic uncertainties of 13.1\%.
%%%%%%%%%%%%%%%%%%%%%%%%%%%%%%%%%%%%%%%%
\section{Summary}
\label{con}
%%%%%%%%%%%%%%%%%%%%%%%%%%%%%%%%%%%%%%%%%%%%%%%%%%
This article highlights our preliminary findings on the systematic uncertainties arising from the variation in the treatment of initial helium abundance on the inferred stellar parameters.

An in depth study on the treatment of initial helium abundance is being addressed in Nsamba et al. (in prep), including a comprehensive comparison to findings of \cite{Valle} based on synthetic stellar data, as well as assessing if the scatter arising from the differences in the treatment of initial helium abundance in stellar grids still satisfy the  stellar parameter accuracy requirements expected for precise exoplanet characterisation for  the future ESA's PLATO space mission. 

%%%%%%%%%%%%%%%%%%%%%%%%%%%%%%%%%%%%%%%%%%%%%%%
\section{Acknowledgements}
%%%%%%%%%%%%%%%%%%%%%%%%%%%%%%%%%%%%%%%%%%%%%%%%%
This work was supported by Funda\c{c}\~{a}o para a Ci\^{e}ncia e a Tecnologia (FCT, Portugal) through national funds (UID/FIS/04434/2013) and by FEDER through COMPETE2020 (POCI-01-0145-FEDER-007672).
 BN acknowledges support from the project "CIAAUP-21/2019-CTTC, Alexander Humboldt Fundation and travel support from  the workshop ``Dynamics of the Sun \& Stars: Honoring the Life \& Work''.
%%%%%%%%%%%%%%%%%%%%%%%%%%%%%%%%%%%%%%%%%%%%%%%%%%%%

%I have done some changes in spbasic in FUNCTION {format.doi} and FUNCTION {format.eprint}
%I have done some changes in mybib.bib in authors and have a backup in olderbib.bib
%\bibliographystyle{spbasic}
%\bibliography{mybib}

%%%%%%%%%%%%%%%%%%%%%%%% referenc.tex %%%%%%%%%%%%%%%%%%%%%%%%%%%%%%
% sample references
% %
% Use this file as a template for your own input.
%
%%%%%%%%%%%%%%%%%%%%%%%% Springer-Verlag %%%%%%%%%%%%%%%%%%%%%%%%%%
%
% BibTeX users please use
% \bibliographystyle{}
% \bibliography{}
%
\newcommand{\mnras}{{MNRAS}}
\newcommand{\aap}{{AAP}}
\newcommand{\apjl}{{APJL}}
\newcommand{\apj}{{APJ}}
\newcommand{\apjs}{{APJS}}
\newcommand{\ssr}{{SSR}}

\end{document}